\documentclass[%
 reprint,
 amsmath,amssymb,
 aps,
]{revtex4-2}
\usepackage[dvipsnames]{xcolor}
\usepackage{graphicx}
\usepackage{dcolumn}
\usepackage{bm}
\usepackage{hyperref}
\usepackage{subfigure}
\usepackage{multirow}
\usepackage{calligra}

\usepackage{tikz} 
\usetikzlibrary{arrows,scopes} 

\newcommand{\rc}{%
\resizebox{!}{1.25ex}{%
    \begin{tikzpicture}[>=round cap]
        \clip (0.09em,-0.05ex) rectangle (0.61em,0.81ex);
        \draw [line width=.11ex, <->, rounded corners=0.13ex] (0.1em,0.1ex) .. controls (0.24em,0.4ex) .. (0.35em,0.8ex) .. controls (0.29em,0.725ex) .. (0.25em,0.6ex) .. controls (0.7em,0.8ex) and (0.08em,-0.4ex) .. (0.55em,0.25ex);
    \end{tikzpicture}%
}%
}

\usepackage[compact]{titlesec}         
\begin{document}
\hbadness=99999
\preprint{APS/123-QED}

\title{Exploring velocity dispersion anisotropy in a dark matter dominated ultra-diffuse galaxy with modified gravity models}

\author{Esha Bhatia}
\email{b.esha@iitg.ac.in}
\author{Sayan Chakrabarti}
\email{sayan.chakrabarti@iitg.ac.in}
\author{Sovan Chakraborty}
\email{sovan@iitg.ac.in}
\affiliation{Department of Physics, Indian Institute of Technology, Guwahati 781039, India}


\date{\today}
\begin{abstract}
 
The kinematics of the ultra-diffuse galaxy (UDG) NGC1052-DF44 is primarily influenced by the presence of dark matter (DM). In this paper, we conduct a contrasting kinematic study of DF44 within the alternative modified gravity framework. In comparison to NFW DM, we test three alternative gravity models viz Milgromian dynamics (MOND), characterized by a known acceleration scale, 
a generic $f(R)$ model, assuming an expansion of the Ricci scalar, and a quantum gravity-inspired Renormalization Group correction to General Relativity (RGGR), which involves the running of the gravitational coupling parameter $G$ with the Universe's energy scale. For each gravity model, we evaluate the velocity dispersion (VD) of the galaxy beyond the conventional radial isotropic assumption and extend to two anisotropy scenarios, i.e., constant and Osipkov-Merritt. Our results show that all three gravity models can provide consistent fits to the observed VD of DF44; however, only MOND and RGGR remain competitive with NFW DM. Interestingly, the constant anisotropy scenario in all the models is also found to be competitive with the complete isotropic assumption. 

\end{abstract}

\maketitle

\section{\label{sec:level1}Introduction}
General Relativity (GR) 
  is well-tested within our solar system scale as well as in astrophysical and cosmological scales. In recent years, two different experiments have validated GR: the observation of Gravitational Waves (GW) \cite{LIGOScientific:2016aoc}, as well as the image of the black hole at the center of the M87 galaxy by EHT collaboration \cite{EventHorizonTelescope:2019dse}. 
  However, on the scales of galaxies GR faces the crucial missing mass problem, i.e., there is a mismatch between the dynamical mass required to keep the gravitational system stable and the baryonic mass contained within, as can be visualized by the Rotation Curve (RC) \cite{1939LicOB..19...41B, Rubin:1978kmz}. The literature suggests two alternative scenarios to explain this missing mass problem. The first possibility is the presence of a large amount of weakly interacting, non-luminous DM in addition to the baryonic mass present within the galaxy \cite{Bertone:2016nfn, Strigari:2012acq, Perivolaropoulos:2021jda, Salucci:2018hqu}. The second is a modification of the Einstein-Hilbert (EH) action of gravity. This may lead to several different types of modified theories of gravity, such as the addition of scalar, tensor, vector fields \cite{Mannheim:2005bfa, Nojiri:2006ri, Clifton:2011jh, Nojiri:2017ncd}, or higher order functional of Ricci scalar \cite{Capozziello:2011et} to name a few. The study of Low Surface Brightness (LSB) galaxies has thus become an important tool for analyzing alternative theories of gravity because of the DM dominance in understanding their kinematics. One subclass of LSB is UDGs having central surface brightness $24.5$ g~mag~ arcsec$^{-2}$, and an effective radius larger than $1.5~kpc$ \cite{vanDokkum:2014cea, Lelli:2024bgp}. The DM-deficit nature of UDGs such as NGC1052-DF2 and NGC1052-DF4 opened a new avenue to compare DM and modified gravity models.\\
\newline
In particular, the Dragonfly Telescope Array (DTA) \cite{Abraham:2014lfa} had observed a large number of UDGs within the Coma cluster \cite{vanDokkum:2014cea, Yozin2015TheQA}. The distinct feature of such UDGs is the presence of numerous globular clusters (GC) within \cite{beasley2016globular}. The observed GCs are approximately $5-7$ times higher than other galaxies whose stellar luminosity is comparatively brighter \cite{van2019spatially, beasley2016globular, peng2016rich}. The formation history of the UDGs is still an open question. Several mechanisms, including tidal stripping \cite{Carleton2018TheFO, Sales:2019iwl}, ``gas stripping" due to the presence of host galaxies near the UDGs \cite{Fujita:2003zj} and galaxy harassment \cite{Abraham:2014lfa} have been proposed. Due to the large globular cluster and the kinematic stability of UDGs within the cluster, such galaxies are expected to be enclosed within a large spheroidal DM halo \cite{peng2016rich}. However, contrary to expectations, two different spectra of galaxies are observed in the Universe. On one end, several examples of such DM-dominated LSBs, such as NGC1052-DF44 \cite{vanDokkum:2016uwg}, VCC1287 \cite{beasley2016overmassive} and Dragonfly 17 \cite{peng2016rich} has been observed. In particular, the LSB galaxy NGC1052-DF44 \cite{vanDokkum:2016uwg} is found to be the largest, within the Coma cluster, showing $98\%$ of its mass to be composed of DM. On the other end, there are galaxies such as NGC1052-DF2 \cite{vanDokkum:2018vup, Laudato:2022vmq, Aditya:2024wdw}, NGC1052-DF4 \cite{van_Dokkum_2019}, AGC114905 \cite{PinaMancera:2021wpc} which, contrary to expected norm, show almost no presence of DM within.\\ 
\newline
In the current study, we are interested in looking into the kinematics of the DM dominated UDG, NGC1052-DF44 \cite{vanDokkum:2016uwg} in light of three alternative gravity models. In contrast to the previous works studying NGC1052-DF2 and NGC1052-DF4 \cite{Bhatia:2023pts, Islam:2019szu}, the dynamics of DF44 is highly dependent on the presence of DM. The radial variation of the measured VD turns out to be as high as $41^{+8}_{-8}$ km/s$^{-1}$ at $5.1$ kpc \cite{van2019spatially}. Thus, the estimated mass-to-light ratio at half-light radius ($4.6^{+0.2}_{-0.2}$ kpc) is 48$^{+21}_{-14}$ M$_{\odot}$/L$_{\odot}$. The ratio suggests a discrepancy between the dynamical and luminous masses present within the galaxy.
The phenomenological study of the DF44 galaxy in the context of the DM model can successfully explain its kinematics. More precisely, the study with a generalized Navarro Frenk White (g-NFW) \cite{1990ApJ...356..359H} and DiCintio \cite{DiCintio:2014xia} density profile can explain the kinematics of DF44 satisfactorily \cite{vanDokkum2019SpatiallyRS}. On the contrary, the kinematics of DF44 is also found to be consistent when probed in the context of alternative gravity models \cite{Islam:2019szu, Freundlich2021ProbingTR, Laudato:2022drz}.  \\
\newline
In this paper, we update the dynamics of UDG in light of three alternative gravity models by utilizing a more robust analysis to compute VD. In particular, we analyze the evolution of VD beyond the radial isotropic scenario, i.e., anisotropic velocity dispersion. For the first case, we look into a well-known gravity model, Modified Newtonian Dynamics (MOND) \cite{1983ApJ...270..371M, Bruneton:2007si, Famaey:2011kh} to study the kinematics of DF44. The analysis in the context of standard MOND, with the UDG mass solely responsible for dynamics, reveals excellent statistical fit, in line with the existing literature \cite{Islam:2019szu}. 
Alternatively, the external field effect (EFE) of the Coma cluster where the galaxy DF44 is embedded may impact the kinematics of the galaxy in the case of MOND. However, \cite{Freundlich2021ProbingTR} showed that incorporating the external field effect of the Coma cluster fails to explain the observed kinematics of the galaxy. 
A similar conclusion was derived for $10$ other UDGs within the cluster. The proposal to resolve the inconsistency of EFE with DF44 includes a screening mechanism, higher mass-to-light ratio, inconsistency in the distance measurement, tidal disruption, etc. \cite{Freundlich2021ProbingTR}. Among the different scenarios, it was shown that an out-of-equilibrium radial infall of DF44 in the Coma cluster may give rise to an observed higher VD or a suppressed EFE for the UDG \cite{Nagesh2024SimulationsOC}.\\
\newline
However, the majority of the analysis on UDG kinematics is in the context of the radial evolution of the VD, i.e., following the isotropic model. In our analysis, we focus on the impact of deviation from this conventional VD evolution. This anisotropy aspect of the problem becomes important for statistical testing of the alternative gravity models. In this regard, we consider the standard MOND as our reference while comparing alternative gravity models. We analyze the three different choices for the anisotropy to explain the observed DF44 VD. We find that a tangential anisotropic motion of the objects is preferred for DF44. Among the three scenarios, the radial anisotropic choice is the least preferred and is quantified using Bayesian Information Criteria (BIC). \\
\newline
Corresponding to MOND, the acceleration scale parameter is constrained from the prior observations \cite{Gentile:2010xt, Famaey:2005fd}. Hence, the study only includes mass modeling parameters to be fitted. Based on the technique applied for the phenomenology of standard isolated MOND study, we look into two distinctive alternative gravity models. One model assumes a generic functional form $f(R)$ in contrast to the Ricci scalar ($R$) in the EH action of gravity. Rather than defining a particular form, this model assumes a general expansion of $R$ about a flat background \cite{Clifton:2011jh,will_1993}. The choice of the particular $f(R)$ model involves constraining two model parameters, which include coupling and scale radius. On the contrary, the other model is the RGGR \cite{Rodrigues:2009vf, Rodrigues:2012qm} that studies the energy scale dependence of the coupling parameter of the theory. On the astrophysical scales, the major contribution to the kinematics arises from the variation of the gravitational coupling parameter. The solution to the potential valid on the galactic scale is dependent on the potential energy of the system and a mass-dependent free parameter. \\
\newline
The paper is organized as follows. Section \ref{sec:vd} discusses the analytical method to study the kinematics of UDG and mass distribution of NGC1052-DF44. Section \ref{sec:model} discusses the three alternative gravity models we look into, to explain the dynamics of DF44.
The methodology employed to constrain the model parameter is discussed in Section \ref{methodology}. Lastly, we discuss the obtained results and the conclusion of our study in Section \ref{res} and \ref{conclude}, respectively.


\section{\label{sec:vd}Velocity dispersion}


 The velocity dispersion ($\sigma$) is a measure of the net galactic kinematics.  Observationally, VD is measured from the broadening of the spectral lines as objects move within the galaxy \cite{2008gady.book.....B}. The radial evolution of the VD is modeled using the Jeans equation, which for a spherically symmetric mass distribution \cite{2008gady.book.....B} is
\begin{equation}\label{sigma}
   \frac{1}{\rho(\rc)} \frac{\partial (\rho(\rc)\sigma^2(\rc))}{\partial \rc}+\frac{\xi(\rc)}{\rc}~\sigma^2(\rc)=\frac{\partial \phi(\rc)}{\partial \rc},
\end{equation}
where $\rho(\rc)$ is the mass density distribution, and $\phi(\rc)$ is the gravitational potential of the galaxy. The deviation from radial isotropy is measured in terms of the anisotropy parameter $\xi(\rc)$, defined as 
\begin{equation}\label{xir}
    \xi(\rc)=1-\frac{\sigma^2_{\theta}(\rc)}{\sigma^2_r(\rc)}.
\end{equation}
Here, $\sigma_r(\rc)$ and $\sigma_{\theta}(\rc)$ are the radial and tangential components of the VD, respectively. Depending on the sign of $\xi(\rc)$, the motion of the object within the galaxy can be either radially ($\xi\geq 0$ ) or tangentially ($\xi\leq 0$) dominated. However, for any isotropic motion of objects in the galaxy, the radial and tangential components are equal, i.e., $\xi=0$.\\
\newline
The right-hand side of the Jeans equation involving the gravitational potential of the galaxy $\phi(\rc)$ can be estimated in the context of Newtonian gravity and for the choice of spherically symmetric mass distribution $\rho(\rc)$,
\begin{equation}\label{npot}
    \frac{\partial \phi(\rc)}{\partial \rc}=\frac{GM_{N}(\rc)}{\rc^2}=\frac{G}{\rc^2}\int_0^{\rc} 4\pi \rho(\rc')\rc'^2 d\rc'
\end{equation}
where $G$ is the Newton's gravitational constant and $M_N$ is the mass within the radius $\rc$. Thus, for a choice of $\xi(\rc)$,  solving Eq.\ref{sigma}, one can obtain $\sigma(\rc)$ given the $\rho(\rc)$ for the galaxy. However, the astrophysical observations do not measure the VD, rather the projection of the radial component of the VD on the line of sight ($\sigma_{LOS}$) is measured. The $\sigma_{LOS}$ from a 2-D projected distance $r$ to the point of observation is expressed as \cite{2008gady.book.....B}  
\begin{align}\label{losvd}
    \sigma_{LOS}^2(r)=\frac{2}{I(r)} \left( \int_r^{\infty} d\rc~\frac{\rc~\rho(\rc)~\sigma^2(\rc)}{\sqrt{\rc^2-r^2}}-\right.\nonumber \\
   \left. r^2\int_r^{\infty} d\rc~\xi(\rc) \frac{\rho(\rc)\sigma^2(\rc)}{r\sqrt{\rc^2-r^2}}\right),
\end{align}

where $I(r)$ is the surface density of the galaxy probed.\\

In the case of alternative gravity models, the potential, in addition to Newtonian contribution (Eq.\ref{npot}), will have an extra component depending on the choice of the gravity model. These components, in total, will result in the net dynamics of the galaxy. 

\paragraph*{{\it{\textbf{Mass model for Ultra-diffuse galaxy (DF44):}}}}
\hfill\break

The kinematics of the galaxy require information about the density distribution of the baryonic contents. In this regard, the computation of Eq.(\ref{losvd}) for a given gravity model has been shown to have a large time complexity. Therefore, to improve the computing time, a reduced analytical form for LOS VD, numerically equivalent to Eq.\ref{losvd} is used \cite{Mamon:2004xk} 
\begin{equation}\label{redlos}
    \sigma_{LOS}^2(r)=\frac{2G}{I(r)}\int_r^{\infty} d\rc~\mathcal{K}\left(\frac{\rc}{r}, \frac{r_a}{r}\right)~j(\rc)\frac{M(\rc)}{\rc}
\end{equation}
where $\mathcal{K}$ is a kernel function, $j(\rc)$ is the projected luminosity density and $M(R)$ is the dynamical mass contained within the galaxy. For the case of GR without DM, the mass function is given by Eq.\ref{npot}. Similarly, for the alternative gravity models, a mass function can be determined from the additional components added to the Newtonian potential. 

\begin{figure*}

        \centering
         \includegraphics[width=0.78\linewidth,height=0.6\linewidth]{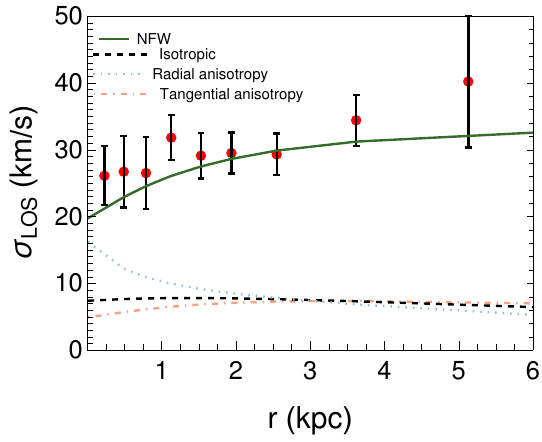}
 
    \caption{\label{fig:newt} The VD for DF44 when the underlying gravity is Newtonian or with an NFW DM halo. The plot represents the Newtonian kinematics of the DF44 galaxy for three different choices of the anisotropic profile. The black dashed, blue dotted, and orange dotted-dashed line highlights the $\xi=0$ (isotropic), $\xi=0.6$ (radial), and $\xi=-0.6$ (tangential) scenarios, respectively. Alternatively, the plot also shows the case for a DM profile represented by a solid green line, which shows a consistent fit for the observational data. The density profile for the DM halo is NFW with $M_{200}=0.70\times 10^{11}~M_{\odot}$.}  
\end{figure*}

The projected luminosity density $j(\rc)$ for a commonly assumed Sersic profile is given as \cite{Neto:1999gx,1997A&A...321..111P}
\begin{equation}
    j(\rc)=j_0~\left(\frac{\rc}{a_s}\right)^{-p_n}\exp\left(-\frac{\rc}{a_s}\right)^{1/n},
\end{equation}
where $a_s=\frac{r_{eff}}{\beta^n}$, $p_n \simeq 1-\frac{0.6097}{n}+\frac{0.05463}{n^2}$ \cite{Neto:1999gx} and $j_0=\frac{L_{tot}}{4\pi n\Gamma[(3-p_n)n] a_s^3}$. The mass modeling estimate for DF44 gives the Sersic parameters, $n=0.94$ and $r_{eff}=4.7 $kpc \cite{vanDokkum:2016uwg}. Also, the total luminosity of the stellar mass contained within the fiducial distance of $100$ Mpc of DF44 comes out to be $L_{tot}=2.33 \times 10^8 L_{\odot}$. Given $L_{tot}$, the central surface brightness $I_0$ in the Sersic profile is calculated from the integrated total luminosity profile, i.e., $2\pi\int_0^{\infty}\rc~I(\rc)~d\rc =L_{tot}$.   \\
The mass density ($\rho(\rc)$) in the context of $M(\rc)$ is also obtained by scaling the projected luminosity density with the constant mass-to-light ratio ($\gamma_{*}$) of the stellar system, i.e.,
\begin{equation}\label{density}
    \rho(\rc)=\gamma_* j(\rc).
\end{equation}
For our statistical analysis, $\gamma_{*}$ is treated as a free parameter having no radial dependence. \\
Finally, the kernel function in Eq.(\ref{redlos}) depends on the nature of the anisotropy parameter. We study the galactic kinematics for three different criteria: $\xi=0$, $\xi$ being constant other than zero, and a radially dependent $\xi$. 
Regarding the radially dependent anisotropy profile, we consider the Osikpov-Merritt model $\xi(\rc)=\frac{\rc^2}{\rc^2+r_a^2}$ \cite{Sotnikova:2008ej,mamon2013mamposst},
where $r_a$ defines the scale radius of the anisotropy profile. In this case, the kernel function is expressed as \cite{mamon2013mamposst},
\begin{align}\label{eq.rxi}
    \mathcal{K}(u,u_a)=\frac{u_a^2+1/2}{(u_a^2+1)^{3/2}}\left(\frac{u^2+u_a^2}{u}\right)\tan^{-1}\sqrt{\frac{u^2-1}{u_a^2+1}}-.\nonumber \\\frac{1/2}{u_a^2+1}\sqrt{1-\frac{1}{u^2}}
\end{align}
where $u=\rc/r$ and $u_a=r_a/r$ as defined in Eq.\ref{redlos}.
The simplest assumption being the isotropic motion of the objects in the galaxy, i.e., $\xi=0$. We also treat $\xi$ as a free parameter that ranges for both the positive (radial) and negative (tangential) anisotropy.\\ 
For the case where $\xi$ is treated as a constant, the kernel function $\mathcal{K}$ in Eq.(\ref{redlos}) is written as
\begin{align}
    \mathcal{K}(u)=\frac{1}{2}~u^{2\xi-1}\left[\sqrt{\pi}\frac{\Gamma(\xi-1/2)}{\Gamma(\xi)}+\xi\right.\nonumber\\\mathcal{B}\left(\frac{1}{u^2},\xi+1/2,1/2\right)
  \left.  -\xi\mathcal{B}\left(\frac{1}{u^2},\xi-1/2,1/2\right)\right],
\end{align}
here $\mathcal{B} (x,a,b)$ is the incomplete Beta function.

Now, we assume the scenario where the dynamics of DF44 can be explained using Newtonian gravity. The LOS VD in such a framework for different anisotropy models is shown in Fig \ref{fig:newt}. The black dashed line in the plot shows an isotropic model, i.e., $\xi=0$ with $\gamma_*=1$. Similarly, the plot also represents the case where the underlying motion in the Newtonian background has a constant anisotropy. The blue dotted line in the plot is a case for radial anisotropy, i.e., $\xi=0.6$, and tangential anisotropy with $\xi=-0.6$ is shown via the orange dotted-dashed line. Indeed, the plot clearly shows that Newtonian gravity alone cannot explain the observational VD for DF44, opening up a scenario for alternative gravity or DM. 

The plot additionally refers to a most commonly studied scenario assuming a DM halo contributing to the Newtonian dynamics.  For this, we look into a well-studied cuspy profile, i.e., NFW \cite{Navarro:1995iw}, assuming the following DM density distribution,
\begin{equation}
    \rho_{NFW} (\rc)=\frac{\rho_s}{\frac{\rc}{r_s}\left(1+\frac{\rc}{r_s}\right)^2},
\end{equation}
where $\rho_s$ and $r_s$ are the characteristic density and radius, respectively. Both the model parameters ($\rho_s$, $r_s$) can be correlated to the concentration parameter ($c=r_{200}/r_s$) and virial mass ($M_{200}$). Using the similar approach as specified in \cite{vanDokkum2019SpatiallyRS} to model DM halo and keeping $\gamma_*$ as an additional free parameter, we obtain the constraints on $\xi$ and $M_{200}$ to be $-0.8$ and $3.98\times 10^{10}$ $M_{\odot}$ ($c$ fixed from $c-M_{200}$ relation \cite{Diemer:2014gba}). The constrained parameters ($\xi$, $M_{200}$) are of similar order as reported in \cite{vanDokkum2019SpatiallyRS}. The radial variation of the VD in the DM framework is shown via a green solid line in Fig.\ref{fig:newt}. The evaluated $\chi^2_{red}$ for the particular DM case is $0.75$. The green dashed line in Fig.\ref{fig:newt} representing the DM case shows that NFW is a consistent choice for the observational VD of DF44.\\
\newline
Thus, DF44 becomes the perfect system to check the alternative scenario of the gravity models in comparison to the DM. Thus, in the following, we study and analyze the consistency of the selected three alternative gravity models, i.e., MOND, $f(R)$, and RGGR, to study the kinematics of the DF44 galaxy. In addition to the gravity model parameters, we have model parameters describing the mass content, such as $\gamma_*$ and the anisotropy parameter ($\xi$), that are also fitted with the observations.    

\section{\label{sec:model}Alternative gravity models}
The modification to the net potential on galactic scales provides an alternative description to the observed discrepancy between the stellar and the total dynamical mass of a galaxy. In view of this, we look into the kinematics of a specific ultra-diffuse galaxy, viz. DF44, from the perspective of three alternative gravity models. The models we focus on are MOND, $f(R)$, and RGGR. 
The phenomenological study of a DM-dominated galaxy such as DF44 in the presence of modified gravity can provide an alternative scenario to explain the overall kinematics without any need to invoke the DM component and, therefore, provides an alternative approach toward studying UDGs that are DM-dominated.    

\subsection{MOND}

Modified Newtonian dynamics (MOND) \cite{1983ApJ...270..371M, Famaey:2011kh} is a theory originally proposed to explain the galaxy rotation curves without invoking the idea of dark matter. The theory relies upon the modification of Newton's second law. 
As is well known, the galaxy rotation curve data points towards the existence of more matter in galaxies than what is observed. The behavior of the rotation curve could be explained by the gravitational force experienced by an object in the outer regions of a galaxy decaying more slowly than predicted by Newtonian gravity. MOND is therefore defined by a critical acceleration scale $a_0$ below which there is a modification to Newtonian gravity. The acceleration in the MOND framework is defined as
\begin{equation}
    \mu\left(\frac{a}{a_0}\right)~a=a_N,
\end{equation}
where $a$ is the net acceleration in the modified framework. In the above equation, $\mu\left(\frac{a}{a_0}\right)$ defines an interpolating function such that for the regions where $a>>a_0$, one regains the correct Newtonian limit. Similarly, in cases where $a<<a_0$, the acceleration ($a$) becomes $\sqrt{a_0a_N}$ \cite{1983ApJ...270..371M}. The best-fit value to the global scale parameter obtained from the fit to the RC of spiral galaxies gives $a_0=1.14 \times 10^{-8}$ cm/s$^2$. For our analysis, we assume a standard form for the interpolating function, which is defined as
\begin{equation}
    \mu(x)=\frac{x}{\sqrt{1+x^2}}.
\end{equation}
Thus, for a scenario where acceleration ($a$) within the galaxy is large such that the external effects can be neglected, the acceleration for the system becomes \cite{1983ApJ...270..371M}
\begin{equation}\label{eq.mondacc}
    a(r)=\frac{GM_{MOND}}{r^2}=\frac{a_N}{\sqrt{2}}\left(1+\left(1+\left(\frac{2a_0}{a_N}\right)^2\right)^{1/2}\right)^{1/2},
\end{equation}
here $a_N$ is the Newtonian acceleration $(\propto 1/r^2)$ .
We need to determine the modified mass function associated with the alternative gravity model to analyze the behavior of LOS VD in an alternative gravity framework as defined in Eq.\ref{losvd}. The mass contribution associated with the MOND framework ($M_{MOND}(r)$) has an additional contribution dependent on the acceleration scale and can be evaluated from Eq.\ref{eq.mondacc}. Substituting $M_{MOND}(r)$ in Eq.\ref{redlos} evaluates the radial line of sight variation of VD in an alternative framework. Although MOND has no free parameters in the model, our analysis includes a mass-to-light ratio $\gamma_*$ and anisotropy parameter that we aim to constrain statistically. We compare three different anisotropic models associated with the analytical VD profile as discussed in Sec.\ref{sec:vd}. The previous literature of MOND that studies the kinematics of DF44 \cite{Islam:2019szu} assumes a Sersic-like density model scaled by a mass-to-light ratio to model the UDG. In our work, we give a more robust analysis by utilizing an equivalent reduced form for LOS VD and studying the favorability of three different anisotropic profiles, as discussed above. 

\subsection{\texorpdfstring{$f(R)$}{TEXT}  gravity}
The $f(R)$ gravity model replaces the Ricci scalar ($R$) in the Einstein-Hilbert (E-H) action with a generalized function of $R$, i.e. $f(R)$ \cite{Clifton:2011jh,will_1993}. However, it is to be noted that the stability criteria puts a constraint on the choice of the functional form of $f(R)$ \cite{Dolgov:2003px, Woodard:2006nt}. Following \cite{Napolitano:2012fp, Lubini:2011pc}, we consider a model with a general Taylor expansion of the functional $f(R)$ form about the Minkowskian background, i.e.,
\begin{equation}
    f(R)\simeq \sum_{i=0}^{\infty} \frac{f_i(0) R^i}{i!},
\end{equation}
where $f_i$-s are the coefficients associated with the $i^{\rm{th}}$ power of Ricci scalar in the expansion. The first term of the series is a constant; hence, it is dropped from the analysis. The solution for the potential in the weak-field limit results in
\begin{equation}\label{yuk_fr}
\phi(r)=-\left(\frac{GM}{1+\delta}\right)\frac{1+\delta e^{-r/\lambda}}{r},
\end{equation}
where $\delta$ is the coupling constant, determining the nature of the additional force arising from the exponential term. As observed from Eq.(\ref{yuk_fr}), the Newtonian potential is rescaled by a factor of $1/(1+\delta)$. Substituting $\delta=0$ in the above equation gives back the Newtonian potential. The coupling parameter of the model, i.e., $\delta$, is constrained to lie within the range $(-1,0)$ \cite{Napolitano:2012fp}. Additionally, $\lambda$ is the scale length, which is the characteristic of the size of the galaxy. As the size and mass of different galaxies differ, the scale parameter $\lambda$ varies with the individual case. However, the force with which the additional Yukawa term couples with the baryonic matter $\delta$ is expected to remain the same. A comparison of the observational VD for DF44 with the analytical model in the $f(R)$ scenario helps to put constraints on the free parameters.\\
\newline
We look into the modified kinematics of the DF44 galaxy in the $f(R)$ framework, similar to the case of MOND. The modified kinematics in the presence of an alternative gravity model is evaluated by substituting the effective mass of the system in the presence of modified gravity model $M_{mog}$ in Eq.\ref{redlos}. However, unlike MOND, the $f(R)$ model has two free parameters, $\delta$ and $\lambda$, which need to be constrained. Additional parameters to be constrained from observations are mass modeling parameters such as $\gamma_*$ and an anisotropy parameter (dependent on the profile discussed in Sec.\ref{sec:vd}).

\subsection{RGGR gravity}
The third model we study to analyze DF44 is inspired by quantum gravity, known as the Renormalization Group correction to General Relativity (RGGR). RGGR studies the running of coupling parameters present in the action of gravity on the astrophysical scales. The RG theory states that the variation of any coupling parameter $g$ with the energy scale $\mu$ at which the model is probed is defined using a $\beta$ function, which takes the general form $\beta=\mu\frac{dg}{d\mu}$.\\
\newline 
Therefore, knowledge of the beta function provides information about the coupling function $g(\mu)$. However, this approach works well in the case
of field theories in flat space-time. The case of gravity needs to be dealt with more carefully. The theory of gravity arising out of the E-H action is known to be non-renormalizable. However, some steps can
be taken towards renormalizing GR by treating it as a field
theory in curved space-time. In the renormalization approach, the gravitational
coupling $G$, which is constant in the far infrared (IR) regime, is
assumed to vary with the energy scale $\mu$. Therefore, to study the quantum effects, it is necessary to determine the RG flow equation for the gravitational constant. 
Motivated by \cite{Fabris:2012wg, Reuter:2004nx}, we assume the $\beta$ function for $G$ to have the form
\begin{equation}
    \mu \frac{dG^{-1}}{d\mu}=2\nu\frac{M_{Planck}}{c^2\hbar}=2\nu G_0^{-1}.
\end{equation}
The above equation provides the solution of the coupling parameter $G$ as a function of the energy scale ($\mu$) 
\begin{equation}\label{energy}
    G(\mu)=\frac{G_0}{1+\nu ln \frac{\mu^2}{\mu_0^2}},
\end{equation}
where $G_0$ is the bare coupling value of gravitational constant as measured in the solar system and $\mu_0$ is the reference energy scale defined such that $G(\mu_0)=G_0$. Also, $\nu$ is a phenomenological parameter, which measures the strength of the coupling \cite{Rodrigues:2009vf}. Thus, the E-H action of gravity, in addition to the Ricci scalar, has a scalar field $G$, which follows the RG equation given in Eq.(\ref{energy}). The previous studies \cite{Rodrigues:2012qm} have shown that a small variation in $\nu$, approximately within the range of $10^{-7}$, significantly impacts the dynamics at the astrophysical scales. 
For the astrophysical scales $\mu$ given in Eq.\ref{energy} can be related to the gravitational potential with a form
\begin{equation}\label{energy_mu}
    \frac{\mu}{\mu_0}=\left(\frac{\phi_N}{\phi_0}\right)^{\alpha},
\end{equation}
where $\phi_0$ is the measured potential on the scales where $G=G_0$. The parameter $\alpha$ in Eq.\ref{energy_mu} has a linear relation with the mass of the galaxy. Solving the equation of motion on the weak-field limit gives the circular velocity written as \cite{Rodrigues:2012wk, Rodrigues:2014xka}
\begin{equation}\label{rgr_v}
    a_{RGGR}(r)=\frac{GM_{RGGR}(r)}{r^2}=a_N(r)\left(1-\frac{c^2\bar{\nu}}{\phi_N(r)}\right),
\end{equation}
$a_N(r)$ and $\phi_N(r)$ in the above equation are the Newtonian acceleration and potential contribution, respectively. Also, the gravity model is constrained by two phenomenological parameters, i.e., $\nu$ and $\alpha$, which can be coupled into a single component, i.e., $\bar{\nu}=\nu\alpha$. In the above Eq.\ref{rgr_v}, $\phi_N(r)$ is the Newtonian potential, which in isolated cases corresponds to the potential energy of the galaxy alone. However, DF44 is embedded in the Coma cluster, which might influence the potential energy of the DF44 galaxy. The effect is similar to the violation of the Strong Equivalence Principle (SEP) in MOND. \\
\newline
All three scenarios for modified gravity models stated in this section reduce to Newtonian gravity at a certain limit. The dependence of the alternative model on the baryonic mass component of a galaxy necessitates the modeling of the galaxy under consideration, which incorporates the free parameters ($\gamma_*$ and $\xi$). On top of the mass-model parameters, depending on the gravity model, we have additional free parameters that need to be constrained from the observational VD data for DF44. Our analysis uses a Bayesian technique to uniquely scan the parameter space for each gravity model.

\begin{figure*}[ht!]

        \centering
         \includegraphics[width=0.78\linewidth,height=0.6\linewidth]{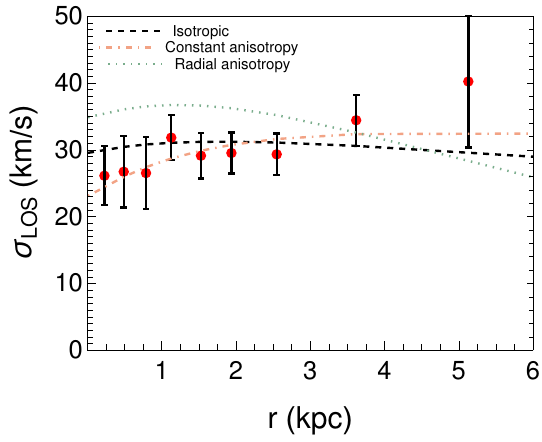}
 
    \caption{\label{fig:mond} The analytical VD radial profile assumes that the underlying gravity is MOND. The dashed black line shows the VD profile for an isotropic $\xi=0$ model in the MOND scenario. The orange dotted-dashed line represents the case for a constant anisotropy, and the green dotted line shows the VD for a radial anisotropy profile as discussed in Eq.\ref{xir}.}  
\end{figure*}

\section{\label{methodology}Methodology}
The constraint on the model parameters for a given gravity model when compared with the observational VD for DF44 is computed via Markov Chain Monte Carlo (MCMC) sampler \cite{ForemanMackey2012emceeTM}. The sampler scans the posterior distribution of the parameter space around some seed value and compares it with the observational data. In our analysis, we aim to run a sufficient number of chains such that the parameters converge toward a value that satisfactorily fits the observations. The posterior probability distribution for the Bayesian model is proportional to the product of likelihood and priors on the parameters. Assuming that the errors in the observations follow a Gaussian distribution, the likelihood is defined as
\begin{multline}\label{likeli}
\mathcal{L}(\bm{\theta})=(2\pi)^{(-N/2)}\left\{\prod_{i=1}^{N} \sigma_{err}(r_i)^{-1} \right\} \times \\ exp\left\{-\frac{1}{2} \sum_{i=1}^{N} \left(\frac{\sigma_{obs}(r_i)-\sigma_{LOS}(r_i,\bm{\theta})}{\sigma_{err}(r_i)} \right)^2\right\},
\end{multline}
where $N$ is the number of observational datapoints and $\sigma_{err}$ represents the error on the observations for a given distance $r_i$. Also, $\sigma_{obs}(r_i)$ corresponds to the VD observations, and $\sigma_{LOS}(r_i)$ is the analytical LOS VD calculated at $r_i$ for the alternative gravity model as defined in Eq.\ref{redlos}. The LOS VD in the modified gravity framework has free model parameters ($\theta$) corresponding to the mass distribution model ($\gamma_*$, $\xi$) and from the chosen alternative gravity model. In the case of MOND, the scale parameter $a_0$ is fixed from the observations, hence we only study the behavior of mass modeling parameters. The $f(R)$ gravity parameters include the coupling parameter $\delta$ and scale radius $\lambda$. Alternatively, the RGGR model has a single mass-dependent parameter $\bar{\nu}$, phenomenologically constrained from the observations. \\
\newline
\begin{figure*}[t!]

        \centering
         \includegraphics[width=0.78\linewidth,height=0.6\linewidth]{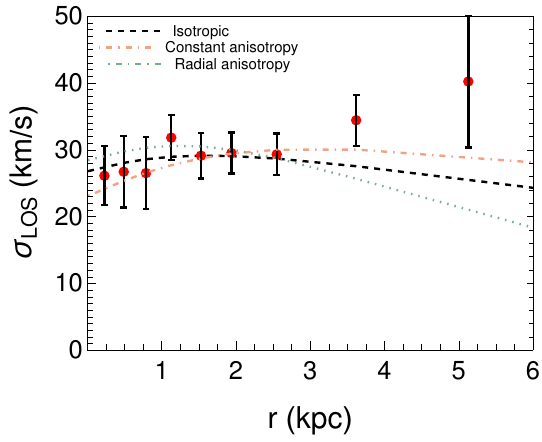}
        \caption{\label{Fig.fRyuk}The radial VD variation for DF44 galaxy under the assumption that the underlying gravity is $f(R)$. The three different anisotropy profiles studied, i.e., $\xi=0$, $\xi=const$, and $\xi(r)$, are shown via black-dashed, orange dotted-dashed, and green dotted lines, respectively. The red dots with the error bar correspond to the observational VD data points for the DF44.}  
\end{figure*}
 This paper aims to test the gravity models for all three scenarios of anisotropy profiles as discussed in Sec.\ref{sec:vd}. The simplest scenario assumes that the tangential and the radial components of the anisotropy parameter in Eq.(\ref{xir}) are equal (isotropic motion), i.e., $\xi(\rc)=0$ and motion of the objects in the galaxy is truly radial. An alternative case assumes that the anisotropy parameter is a constant between ($-\infty,~1$). The negative values of $\xi$ indicate that the kinematics of the object in the galaxy is dominated along the tangential orbit. Alternatively, $\xi=1$ indicates that the orbits of the clusters are completely radial. For the third case, we assume the anisotropy profile to have an Osipkov-Merritt form as given in Eq.\ref{xir}. This radial profile introduces a free parameter, scale radius ($r_a$). Thus, in addition to gravity model parameters and mass to light factor, we have additional parameters coming from the choices of $\xi$ in the definition of VD. 
For our sampler, we assume flat priors on the parameter space that is varied in a wide range. For the scenario where $\xi(\rc)$ is treated as a constant, the parameter varies within a wide range of ($-10$, $1$). This choice for the variation in $\xi$ incorporates the anisotropy parameter's radial and tangential behavior. Similarly, for the third case where $\xi(\rc)$ has a radial dependence, the anisotropy profile is parameterized by scale radius $r_c$ and is varied across the scale of the galaxy. \\
\newline
To compare the three scenarios and infer the preference of one over the other, we take the help of Bayesian Inference Criteria (BIC)\cite{1978AnSta...6..461S}. BIC works on the principle of maximum likelihood and is defined as
\begin{equation}
    BIC=-2\log \mathcal{L}_{max}(\bm{\theta})+2k\log(n);
\end{equation}
where $k$ is the number of free parameters present in each case for a model, and $\mathcal{L}$ is the likelihood function defined in Eq.(\ref{likeli}). To quantify the favorability of the model, we evaluate
\begin{equation}
    \Delta BIC=BIC_2-BIC_1
\end{equation}
A positive difference between the BIC of the two anisotropy scenarios hints towards the preference of the first model over the second. According to the criteria, if the difference is less than $2$, it suggests that both the models are performing equally well, and the result is inconclusive. Alternatively, if the difference is more significant than $2$, it suggests an inclination toward the second model with $\Delta BIC$ between $2-6$, implying a positive inclination, and greater than $6$ is considered a strong inclination toward the second model. Additionally, to ensure the convergence of the sample, the chain runs for a sufficient number of steps, and the acceptance fraction lies between $0.2-0.8$ \cite{ForemanMackey2012emceeTM}.
\begin{figure*}[t!]

        \centering
         \includegraphics[width=0.78\linewidth,height=0.6\linewidth]{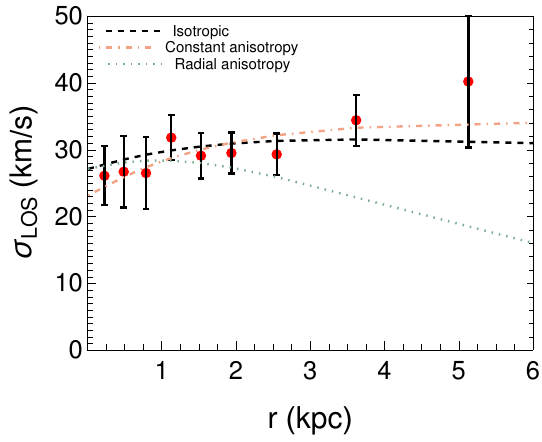}
 
    \caption{\label{fig:rgr} The radial VD obtained for DF44 when the underlying gravity is RGGR. The plot shows the VD modeling for three assumptions of anisotropy. The black dashed line represents the isotropic case; the orange dotted-dashed line is for the constant anisotropy, and the green dotted line represents the radial anisotropy case.}  
\end{figure*}
\section{\label{res}Results}

\subsection{MOND}
Our analysis looks into MOND for different choices of the anisotropy parameter as discussed in Sec.\ref{sec:vd}. For the first case, the VD kinematics is assumed to be isotropic, i.e., $\xi=0$. Thus, the model has a single free parameter $\gamma_*$ that is constrained from the data. For the particular scenario, the best-fit value obtained for $\gamma_*$ is $1.02$. Substituting the best-fit value, a dashed black line in Fig \ref{fig:mond} shows the radial variation of VD in the MOND framework. The red dots with the error bar constitute the observational dataset for DF44 \cite{vanDokkum:2016uwg}. Similarly, the second case assumes $\xi$ as a constant with no radial dependence. Thus, depending on the signature of $\xi$, the dynamics of the galaxy can be radial ($\xi>0$) or tangential ($\xi<0$) in nature. Constraining the two free parameters \{$\gamma_*$, $\xi$\}, we obtain mass-to-light ratio $\gamma_*=1.31$ and $\xi=-0.42$. The negative value of $\xi$ highlights the tangential nature of the anisotropy profile. The variation of VD for the second case obtained by substituting the best-fit value is plotted via an orange dotted-dashed line in Fig \ref{fig:mond}. 
The third case for the radially dependent anisotropic profile introduces the free parameter ($r_a$), determining the scale radius of the anisotropy profile. For this case, the radial VD is shown using a green dotted line in Fig \ref{fig:mond} with the obtained best-fit values, i.e., $\gamma_*=1.02$ and ${r_a}=5.73~kpc$. The summarized constrained parameters obtained for all different choices of anisotropy are also compiled in Table.\ref{Tab:mondpar}.  


\begin{table}[ht]
\begin{tabular}{|*{7}{c|}} 
\hline
$\xi$~&~$\gamma_*$~&~$\chi^2_{red}$&~$BIC$~\\
\hline
$\xi=0$ & $1.02$& $0.64$ &$9.49$\\
\hline
$\xi=const=-0.41$ &$1.31$ & $0.39$ & $11.51$\\
\hline
$\xi(\rc) (r_a=5.67~kpc)$ & $1.00$ &$3.96$& $36.54$\\
\hline
\end{tabular}
\caption{The constrained best-fit model parameters in MOND. The three scenarios in the table represent the different choices for the anisotropy parameter with $\xi=0$, $\xi$=const, and $\xi(r)$ as the isotropic, constant, and Osikpov-Merritt profile. The acceleration scale for the MOND model ($a_0$) is fixed from the observations. The $\chi^2_{red}$ measures the goodness of fit for each case.}
\label{Tab:mondpar}
\end{table}
Table.\ref{Tab:mondpar} also contains the measured $\chi^2_{red}$ inferred from the best-fit MOND parameter for the three choices of the anisotropy model, together with the BIC values. 
From Table.\ref{Tab:mondpar}, we observe that the $\gamma_*$ obtained in all three scenarios are similar and are consistent with the previous analysis done in \cite{Islam:2019szu}.  The difference in BIC between the radial and constant anisotropy case ($\Delta BIC=25.03$) points towards the preference of the latter choice. However, a comparison of the isotropic and best fit constant anisotropy case yielding a small $\Delta BIC=2.1$ hints that the preference of one model over the other is inconclusive.  

\subsection{\texorpdfstring{$f(R)$}{TEXT}  gravity}
The $f(R)$ model we study assumes a general Taylor expansion of the functional form of the Minkowskian background, i.e., $R=0$. The weak-field potential obtained for the model is characterized by two parameters, viz., the coupling parameter $\delta$ and the scale radius $\lambda$. In addition to these alternative parameters of the gravity theory model, we have the usual mass-to-light ratio ($\gamma_*$) and anisotropy parameter $\xi(\rc)$, which are also constrained statistically. The suggested variation of the coupling parameter $\delta$ lies within the range $(-1,0)$, and $\lambda$ is scaled within the size of the galaxy. Additional priors on the parameters $\delta$ and $\lambda$ remain the same as discussed in Sec. \ref{methodology}.
\begin{table}[ht]
\begin{tabular}{|*{7}{c|}} 
\hline
$\xi$~&~$\gamma_*$~&~$\delta$~&~$\lambda (kpc)~$&~$\chi^2_{red}$&~$BIC$~\\
\hline
$\xi=0$ & $1.56$& $-0.89$ & $0.81$& $1.09$ &$19.47$\\
\hline
$\xi=const=-0.17$ &$1.83$ & $-0.90$ & $2.46$ & $0.89$ & $22.05$\\
\hline
$\xi(\rc) (r_a=4.39~kpc)$ & $2.49$ & $-0.81$ & $3.49$ &$2.13$& $28.25$\\
\hline
\end{tabular}
\caption{The model parameters constrained for the $f(R)$ model in case of DF44. In addition to the model parameters, the table contains $\chi^2_{red}$ for the individual case. We additionally measure the BIC to compare the favorability among the different anisotropy models. }
\label{Tab:frpar}
\end{table}
\newline
Similar to the case of MOND, the $f(R)$ model is also treated for three different assumptions of the anisotropy parameters $\xi$. For 
 the first isotropic case  i.e., $\xi=0$, the analysis depends on three free parameters. The best-fit value obtained for the model parameters i.e., \{$\gamma_*$, $\delta$, $\lambda$\} evaluates to \{$1.56$, $-0.89$, $0.81$ kpc\}. For an alternative case where $\xi$ is treated as a constant, the model parameters, i.e., \{$\xi$, $\gamma_*$, $\delta$, $\lambda$\} obtained are \{$-0.17$, $1.83$, $-0.90$, $2.46$ kpc\}. The negative value for the anisotropy profile suggests a preference towards a tangential profile. The third scenario, assuming a generalized radial anisotropy profile, adds the scale-dependent parameter $r_a$. The best-fit values obtained for these parameters \{$r_a$, $\gamma_*$, $\lambda$, $\delta$\} result in \{$4.39$ kpc, $2.49$, $-0.81$, $3.49$ kpc\}. The parameters corresponding to the three different cases discussed above are also tabulated in Table. (\ref{Tab:frpar}). \\
\newline
\begin{figure*}[ht!]
    \includegraphics[width=0.78\linewidth,height=0.6\linewidth]{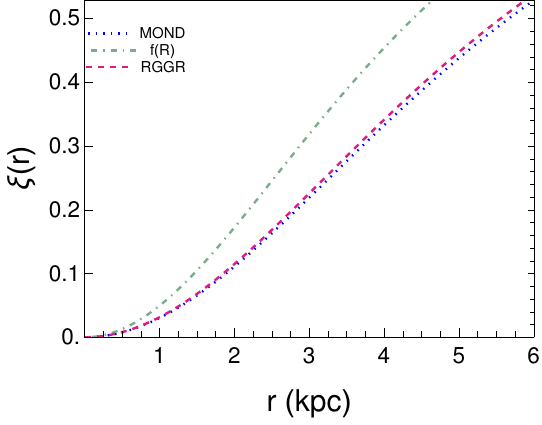}
    \caption{The radial Osikpov-Merritt profile for three modified gravity models. The blue dotted line depicts the anisotropy variation for the MOND model. Similarly, the green dotted-dashed and pink dashed line represents the $f(R)$ and RGGR model, respectively. }
    \label{fig:xir}
\end{figure*}
Additionally, the LOS VD obtained by substituting the best-fit values obtained for each anisotropy  $f(R)$ gravity model is shown in Fig.\ref{Fig.fRyuk}. The plot shows the radial variation of VD in $f(R)$ model with different assumptions for the anisotropy parameter, i.e., $\xi=0$ (left), $\xi=const$ (middle) and $\xi(\rc)$ (right). The dashed black line in the plot shows the VD modeling for the $f(R)$ gravity obtained by substituting the best-fit values corresponding to the isotropic case. Similarly, the case for a constant and radial choice of the anisotropy profile is shown via orange dashed-dotted and green dotted lines, respectively. As seen from Table.\ref{Tab:frpar}, the $\chi^2_{red}$ is smaller for the case where anisotropy is treated to have no radial dependence. 
Note that although the second and third scenarios have a similar number of free parameters, the change in BIC suggests an inclination toward a constant anisotropic case. Additionally, the isotropic and tangential anisotropy modeling behave equally well with the observation that can be quantified from the difference in BIC ($\Delta BIC \sim 2$).

\begin{figure*}[ht!]
    \centering
    \includegraphics[width=0.78\linewidth,height=0.6\linewidth]{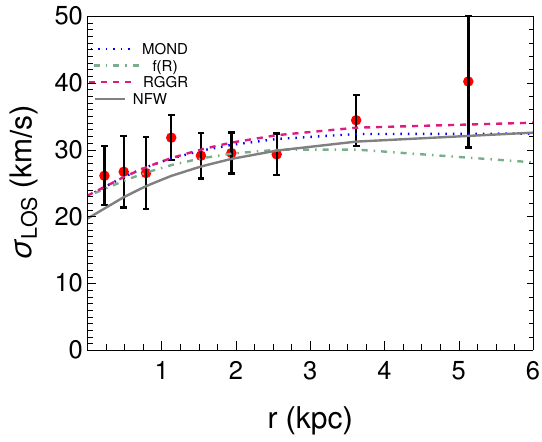}
    \caption{The plot compares the three alternative gravity and NFW DM models for the constant anisotropic case. The blue-dotted, pink-dashed, and green-dotted-dashed lines correspond to the MOND, RGGR, and Yukawa models, respectively. The gray solid line in the plot represents the alternative DM scenario. }
    \label{fig:4_mod}
\end{figure*}
\subsection{RGGR gravity}
The solution to the weak-field limit potential in RGGR introduces a mass-dependent phenomenological parameter $\bar{\nu}$. The previous studies for the RGGR model suggest $\bar{\nu}$ to be in the order of $10^{-7}$ for spiral and elliptical galaxy RC data \cite{Rodrigues:2009vf, Rodrigues:2012qm, Bhatia:2024tak} and $10^{-8}$ for UDG VDs \cite{Bhatia:2023pts}.
Based on these results, for our study, we assume flat priors on the RGGR parameter ranging from [$10^{-8}$-$10^{-6}$]. Additionally, priors on the VD parameters such as ($\gamma_*$, $\xi$) remain similar to those discussed in the previous two alternative gravity models. For our analysis of the DF44 galaxy, we probe the kinematics for two RGGR frameworks, i.e., an isolated scenario and under the influence of external effects, as discussed below.  
\begin{table}[ht]
\begin{tabular}{|*{6}{c|}} 
\hline

$\xi$~&~$\gamma_*$~&~$\bar{\nu}$$\times$$10^{-8}$~&~$\chi^2_{red}$&~$BIC$~\\
\hline
$\xi=0$ & $1.45$& $2.46$ & $0.43$& $11.82$\\
\hline
$\xi=const.=-0.34$ &$1.59$ & $2.75$ & $0.45$ & $15.88$\\
\hline
$\xi(\rc)(r_a=5.56~kpc$) & $1.44$ & $1.79$ & $1.53$ & $22.36$\\
\hline
\end{tabular}
\caption{The best-fit model parameters for the RGGR gravity model. The table contains the constrained values obtained for gravity ($\bar\nu$) and mass-model ($\xi$, $\gamma_*$) parameters. The three scenarios studied include isotropic $\xi=0$, constant ($\xi=const.$), and radial ($\xi(r)$) profile.}
\label{Tab:rgrpar}
\end{table}
For the standard RGGR scenario, the potential energy contribution is comprised of the matter density of the DF44 galaxy alone. Similar to the previous models, we study three different assumptions of the anisotropy parameter. The first case with $\xi=0$ involves two model parameters, i.e., $\bar{\nu}$ and mass-to-light ratio ($\gamma_*$), to be constrained from the observations. The best fit parameters evaluated from our study give $\gamma_*=1.45$ and $\bar{\nu}$ as $2.46\times10^{-8}$. For the second scenario, where the anisotropy parameter $\xi$ is treated as a constant, the estimated values of the three free parameters, i.e., ($\xi$, $\gamma_*$, $\bar{\nu}$) come out as ($-0.34$, $1.59$, $2.75\times10^{-8}$ ). For the third choice of radial anisotropy (eq.\ref{xir}), the best-fit values for \{$\gamma_*$, $\bar{\nu}$, $r_a$\} are $1.44$, $1.79 \times 10^{{-8}}$, and $5.56~kpc$ respectively. Indeed, the constrained $\bar\nu$ is an order lesser than obtained from the study of spiral and elliptical galaxies but is consistent with the study of other UDGs \cite{Bhatia:2023pts}. The best-fit model parameters obtained using the sampler are also listed in Table \ref{Tab:rgrpar}.\\ 
\newline

A comparison of the radial VD profile obtained for standard RGGR scenario from statistical analysis with the observational data is shown in Fig.\ref{fig:rgr}. The three assumptions for the anisotropy profile are shown with a black dashed line corresponding to $\xi=0$, orange dashed-dotted for $\xi=const$ case, and the green dotted line studying the radial behavior $\xi(\rc)$ of the anisotropy parameter. Like the other modified gravity models studied previously, the constant anisotropic case favors a tangential profile. We also report the BIC using the best-fit parameters obtained for the three choices of the anisotropy model. The measured $\Delta BIC$ clearly shows that the radial anisotropic choice is the least favored among the three models. Additionally, between the isotropic and the tangential model, $\Delta BIC$ hints towards a slight favorability of the $\xi=0$ choice rather than the constant anisotropic model.  

To summarize, the study of the kinematics of DF44 in light of three gravity models, i.e., MOND, $f(R)$, and RGGR, shows many important features. While comparing all three models, $\Delta$BIC points towards the fact that the radial dependence choice of $\xi(\rc)$ is the least favorable. Using the best-fit $r_a$ obtained for all three models, the radial variation for the Osikpov-Merritt profile within the size of the galaxy is shown in Fig\ref{fig:xir}.  It is also to be noted that for all three models, when the anisotropy parameter is constant, the best-fit value points towards a negative value, i.e., the tangential behavior. 
\\
\newline

\subsection{DM scenario}
To check the favorability of the alternative gravity models, we compare them with the alternative NFW scenario. As the constant anisotropy case for all three gravity models shows a consistent fit with the DF44 observations, we compare the modified gravity analysis with a similar DM scenario to study the favorability of the models. 
\newline
\begin{table}[ht]
\begin{tabular}{|>{\centering\arraybackslash}p{1.5cm}|>{\centering\arraybackslash}p{1.5cm}|>{\centering\arraybackslash}p{1.5cm}|} 
\hline
Model & $\chi^2_{red}$ & $BIC$ \\
\hline
MOND & $0.39$ & $11.51$ \\
\hline
$f(R)$ & $0.89$ & $22.05$ \\
\hline
RGGR & $0.45$ & $15.88$ \\
\hline
NFW & $0.75$ & $17.69$ \\
\hline
\end{tabular}
\caption{The table summarizes the goodness of fit and BIC for the DM and the alternative gravity models for a constant anisotropic parameter.
}
\label{Tab:grav_nfw}
\end{table}

For the DM halo (see Sec.\ref{sec:model}), we found the best-fit value of the anisotropy parameter to be $-0.8$ and $3.98\times10^{10} M_{\odot}$ for $M_{200}$, with $\chi^2_{red} = 0.75$. The BIC evaluated using these best-fit DM parameters gives $17.69$. The details of the goodness of fit and BIC measured for the three gravity models and the NFW scenarios are also compiled in Table. \ref{Tab:grav_nfw}. The three gravity models, along with the NFW scenarios for the tangential anisotropy case, are shown in Fig.\ref{fig:4_mod}. The gray solid line in the plot represents the case of the NFW DM halo. Similarly, the blue dotted, pink dashed, and green dotted-dashed line represents the MOND, RGGR, and $f(R)$ models, respectively. The modified gravity models shown in Fig.\ref{fig:4_mod} for a constant anisotropic choice highlight that all three models are competing choices for the kinematics of DF44 when compared with the DM scenario. A comparison of $\Delta BIC$ between the MOND scenario having a constant anisotropy with the DM model shows that the former is a preferred choice to explain the VD kinematics for DF44. Alternatively, the measured $\Delta BIC$ with RGGR suggests that both models perform equally well. However, a comparison with the $f(R)$ model shows a slight preference for the NFW DM halo. 

\section{\label{conclude}Conclusion}
In this paper, we look into the kinematics of the DM-dominated ultra-diffuse galaxy (DF44) in the context of alternative gravity models, i.e., MOND, $f(R)$, and RGGR. Such UDGs where the galactic kinematics of VD conventionally fit well with DM are in contrast to the DM deficit UDGs like DF2 and DF4. Thus, the phenomenology and suitability of the modified gravity models in the context of these DM-dominated UDGs in comparison to the DM explanation is most challenging. To make things more demanding, the resulting galactic kinematics is not considered to be the conventional radially isotropic one, i.e., anisotropy in VD is taken into account and probed statistically.\\
\newline
We model the DF44 VD for three scenarios of the anisotropy parameter. MOND is considered the reference model to understand the influence of anisotropy on VD regarding these modified gravity models. MOND is parameterized by a single acceleration parameter, fixed globally from prior observations. In the case of $f(R)$, we assume a general Taylor series expansion of the functional form about the Minkowskian background. This assumption adds a Yukawa-like term to the Newtonian potential constrained by free parameters $\delta$ and $\lambda$. Additionally, we look into the RGGR model, which studies the scale-dependent variation of the gravitational constant. The weak-field potential obtained for the model has a mass-dependent free parameter $\bar{\nu}$ that is constrained from the observations. These gravity models are also compared with the DM NFW scenario. The consistency of a gravity model is statistically quantified by $\chi_{red}^2$ analysis. Additionally, we have evaluated BIC to check if one anisotropy scenario is more favorable than another. \\
\newline
In the case of MOND, a comparison of $\chi^2_{red}$ for the three anisotropy cases shows that a choice of radial profile fits poorly with the observations. This can also be observed from the difference in BIC between the scenarios. It is also noted that for all three gravity models, the $\chi^2_{red}$ is slightly improved or remains similar as one moves from isotropic to a constant anisotropy case. However, the difference in BIC between the two cases ($\xi=0$ and $\xi=const$) shows that the results are inconclusive to favor one model over the other for the MOND and $f(R)$. Similarly, for the RGGR model, the isotropic model has a slight edge over the tangential choice of the anisotropy parameter. For the case of the $f(R)$ model, the $\chi^2_{red}$ for isotropic versus $\xi=const$ case shows that both models fit equally well with the observations of DF44. A small difference of $\Delta BIC$ between the two scenarios shows that the result is inconclusive to favor a certain model.  However, similar to MOND, the Osikpov-Merritt profile is the least favorable among the three choices of anisotropy profile. The RGGR model has a single mass-dependent parameter $\bar{\nu}$. For the isolated RGGR case, the best-fit value obtained for the parameter is an order less than constrained by the spiral or elliptical galaxies. However, in line with the previous claim towards ultra-diffuse galaxies (DF2 \& DF4), the order of $10^{-8}$ is consistent for DF44. Also, similar to the other two gravity models, the $\Delta$BIC measured suggests that the radial anisotropy profile is the least favorable among the three choices for the anisotropy.   \\
Finally, in this analysis, we compare our choice of gravity models with that of an NFW DM halo model. The comparison of BIC evaluated for each case suggests MOND as a preferred choice. However, the DM model has a slight edge over the $f(R)$ model, whereas RGGR performs equally well.  \\
\newline
In this first detailed analysis of the anisotropy aspect, we avoid the complications of EFE, i.e., the effective potential energy contribution in the model from the cluster where the galaxy is embedded. Here, the Coma cluster may have an additional contribution from the external field. The LOS VD solution where the external contribution is incorporated 
has been previously studied for the MOND scenario and fails to explain the VD of DF44 \cite{Freundlich2021ProbingTR}. The results in the LOS VD with EFE remain only comparable to the Newtonian scenario, which is inconsistent with the observed VD for the DF44. Some measures have already been proposed to resolve this issue; in particular, any possible out-of-equilibrium radial infall of DF44 in the Coma cluster may give rise to the observed higher VD or a suppressed EFE for the UDG \cite{Nagesh2024SimulationsOC}.
Similarly, EFE analysis with the other alternative gravity models may also reveal the interplay of the different physical model parameters, anisotropy, and mass distribution. However, this is beyond the scope of this analysis.\\
\newline
In summary, UDGs, such as DF44, are a favorable place to test the signature of gravity models, as they are heavily dominated by the DM. The anisotropy of the radial evolution of the VD can impact fitting quality. We have compiled an in-depth analysis of the DF44 observations in light of three modified gravity models with the anisotropy analysis. All three alternative gravity models provide a consistent picture of the dynamics of DF44. For all three choices of gravity models (MOND, $f(R)$, and isolated RGGR), the constant anisotropic case shows an inclination toward the tangential profile instead of radial.
Additionally, analysis with the Osikpov-Merritt profile shows that the radial nature of the anisotropy profile is less agreeable with the observations compared to the other two cases. We have additionally reported the analysis with an alternative picture where an NFW halo influences the kinematics of DF44. The comparison clearly shows that the choice of alternative gravity models competes with the DM scenario to explain the kinematics of the UDG DF44. 
Future data on more such UDGs will open up the possibility of global analysis accessing stricter statistical testing of these models. 
\acknowledgments
Sovan and Esha would like to acknowledge the support of DST-SERB projects CRG/2021/002961. Sovan also acknowledges the support from DST-SERB research grant number MTR/2021/000540. Sayan  acknowledges the support from the DST-SERB research grant
MTR/2022/000318. In addition Esha acknowledges the informative discussions regarding RGGR with Jonathan Freundlich and Davi C. Rodrigues.
\nocite{*}

\bibliography{apssamp}

\end{document}